\documentclass[a4paper,12pt]{article}
\usepackage{mathrsfs}
\usepackage{amsfonts}
\usepackage{amssymb}
\usepackage{amsthm}
\usepackage{amsmath}
\usepackage{authblk}
\usepackage{url}
\usepackage{color}
\allowdisplaybreaks

\newtheorem{Theorem}{\sc Theorem}%%%[section]

\newtheorem{Lemma}[Theorem]{\sc Lemma}

\newtheorem{Problem}[Theorem]{\sc Problem}

\def\sqr#1#2{{
    \vcenter{
         \vbox{\hrule height.#2pt
               \hbox{\vrule width.#2pt height#1pt \kern#1pt
                     \vrule width.#2pt
               }
               \hrule height.#2pt
         }
    }
}}
\def\square{\mathchoice\sqr84\sqr84\sqr{2.1}3\sqr{1.5}3}

\def\div{\mathop{\rm div}\nolimits}
\def\Div{\mathop{\rm Div}\nolimits}

\def\bar{\overline}
\def\real{\mathbb{R}}

\newcommand{\R}{{\if mm {\rm I}\mkern -3mu{\rm R}\else \leavevmode
\hbox{I}\kern -.17em\hbox{R} \fi}}

\def\lista#1
{{ \itemindent 0.0cm \labelsep .2cm \leftmargin 0.8cm \rightmargin
0.0cm \labelwidth 0.6cm \topsep 0.0mm
\parsep 0.0mm
\itemsep 0.0mm
\begin{list}{}
{ \setlength{\leftmargin}{1.0cm} \setlength{\rightmargin}{0.0cm}
\setlength{\parsep}{.0mm} \setlength{\topsep}{.0mm}
\setlength{\parskip}{.0cm} \setlength{\itemsep}{.0cm} }
{#1}\end{list}} }

\begin{document}

\title{ 
Nonmonotone slip problem for miscible liquids
\thanks{
	\, The project has received funding from the European Union's Horizon 2020 Research and Innovation Programme under the Marie Sk{\l}odowska-Curie grant agreement No. 823731 -- CONMECH.
	It has been supported by the National Science Center of Poland under Maestro Project No. UMO-2012/06/A/ST1/00262, 
	the Qinzhou University Project 
	No. 2018KYQD03, and the International
	Project co-financed by the Ministry of Science and Higher Education of Republic of Poland under Grant No. 3792/GGPJ/H2020/2017/0.
}}

\author{Stanislaw Mig\'orski$^{\,1,2}$\ \ }
\author{
Pawel Szafraniec$\,^{2,3}$ \footnote{\, Corresponding author. E-mail: pawel.szafraniec.wmii@gmail.com 
	(P. Szafraniec).}~}

\affil{   
$^1$ 
Chengdu University of Information Technology\\
College of Applied Mathematics \\ 
Chengdu, 610225, Sichuan Province, P.R. China \\
\smallskip\smallskip
$^2$ Jagiellonian University in Krakow \\
Faculty of Mathematics and Computer Science \\
ul. Lojasiewicza 6, 30-348 Krakow, Poland \\
\smallskip\smallskip
$^3$  University of Agriculture in Krakow \\
Faculty of Production and Power Engineering \\
ul. Balicka 116B, 30-149 Krakow, Poland
}

\date{}

\maketitle
\thispagestyle{empty}
\bigskip

\noindent {\bf Abstract.} \  
In this paper we prove the existence and uniqueness of a solution to the nonstationary two dimensional system of equations describing miscible liquids with nonsmooth, multivalued  
and nonmonotone boundary conditions of subdifferential type. 
We employ the regularized Galerkin method combined with results from the theory of hemivariational inequalities. 

\bigskip

\noindent {\bf Keywords:} 
Navier-Stokes equation; generalized subgradient; 
nonconvex potential; operator inclusion; weak solution.

\vskip 4mm

\noindent {\bf 2010 Mathematics Subject Classification: } 
76D05, 76D03, 35D30, 35Q30.

\newpage

\section{Introduction}\label{Introduction}

In this paper we consider the mathematical model for two dimensional miscible liquids and provide a result on existence and uniqueness of weak solution under a nonmonotone slip boundary condition.
The model is a system of partial differential equations which consists of Navier-Stokes equations with Korteweg stress terms
for the velocity and pressure of the fluid coupled 
with the reaction-diffusion equation for the concentration of the fluid.

Miscibility is the property of substances to fully dissolve 
in each other at any concentration forming a homogeneous solution. This notion is mostly applied to liquids, but applies also to solids and gases. 
Two liquids are miscible if the molecules of the one liquid can mix freely with the molecules of the other liquid forming a uniform blend. For historical reasons the substance less abundant in the mixture is called
a solute, while the most abundant one a solvent.
There is no sharp interface between miscible liquids,
but rather a transition zone. Examples of such phenomenon is the mixing of water and glycerin, and water and ethanol.
The study of miscible liquids is motivated by problems 
in oil recovery, hydrology, polymer blends,
groundwater pollution and filtration~\cite{Allali0,Allali1,AVV,BPVVZ,KOSTIN}.

It was experimentally confirmed that between two miscible liquids there exists a transient capillary phenomena 
since the change of concentration gradients near the transition zone causes capillary forces between two liquids, see~\cite{BPV}. 
For this reason due to the concentration inhomogeneities, 
we need to take into account additional terms in the equation of motion. 
These terms introduced first in the work by Korteweg~\cite{Kor} represent additional volume 
forces in the equations of motion called now
Korteweg stresses.

Results on the unique weak solvability 
of models describing miscible liquids 
can be found~\cite{Allali0,KOSTIN} where the problem 
was studied in two dimensional case, 
in the absence of external and source forces, 
with no subdifferential boundary conditions, 
and in~\cite{Allali1,AVV} who treated the three dimensional 
case with the homogeneous Dirichlet boundary condition 
on the whole boundary.
A result on existence of the global weak solution for a
multiphasic incompressible fluid model with
Korteweg stress can be found in~\cite{CEZ} where the Galerkin method combined with a fixed point argument have been employed.

The remainder of the paper is as follows. 
In Section~\ref{Preliminaries} we recall some preliminary material and the functional setup of the problem. The classical and variational formulations of a model of miscible liquids are described in Sections~\ref{Statement}
and~\ref{variational}, respectively. 
Section~\ref{proof} is devoted to the proof of Theorem~\ref{maintheorem} 
which is the main result of the paper on existence and uniqueness of weak solution to the model.

\section{Notation and preliminaries}\label{Preliminaries}

In this section we introduce notation and recall some preliminary material.

Let $\Omega$ be a bounded open subset of $\real^2$ 
with boundary $\Gamma$ of class $C^2$ 
composed of two disjoint measurable 
parts $\Gamma_0$ and $\Gamma_1$, i.e.,  
$\bar{\Gamma}_0 \cup \bar{\Gamma}_1 = \Gamma$ and  $\Gamma_0\cap \Gamma_1 = \emptyset$ 
with ${\rm meas}\, (\Gamma_0)>0$. 
Given a vector $\xi\in \real^2$ on the boundary $\Gamma$, 
we denote by $\xi_\nu$ and $\xi_{\tau}$ its normal and tangential components, respectively, i.e., 
$\xi_\nu = \xi \cdot \nu$ 
and $\xi_{\tau}=\xi -\xi_\nu\nu$, where $\nu$ denotes the outward normal unit vector to the boundary. 
%%Here $``\cdot"$ denotes the standard 
%%inner product in $\real^2$.
The notation $\mathbb{S}^{2}$ represents the class of second order symmetric $2 \times 2$ tensors. 
The inner products and norms in 
$\mathbb{R}^{2}$ and $\mathbb{S}^{2}$ are denoted by
\begin{eqnarray*}
	&& u\cdot v=u_{i}v_{i},\qquad\|v\|=(v\cdot v)^{1/2}
	\qquad \textrm{for all } \ u=(u_{i}), \ v=(v_{i})\in \mathbb{R}^{2},\\[1mm] 
	&&
	\sigma:\tau=\sigma_{ij}\tau_{ij},~\quad\|\tau\|
	=(\tau:\tau)^{1/2}
	\qquad \textrm{for all } \ \sigma=(\sigma_{ij}), \ \tau=(\tau_{ij})\in
	\mathbb{S}^{2},
\end{eqnarray*}
respectively.
We introduce the following function spaces
\begin{eqnarray*}
&&
E=\{\ v \in H^1(\Omega)^2 \mid v = 0 \ \mbox{on} \ \Gamma_0, \ v_\nu =0 \ \mbox{on} \ \Gamma_1, \ \div{v} = 0 \ \},
\\[2mm]
&&
H=\{\ 
v \in L^2(\Omega)^2 \mid v = 0 \ \mbox{on} \ \Gamma_0, \ v_\nu =0 \ \mbox{on} \ \Gamma_1, \ \div{v} = 0 \ \} , \\[2mm]
&&
V=\{ \ v \in H^2(\Omega) \mid \frac{\partial v}{\partial \nu}=0\ \}.
\end{eqnarray*}

We denote by $X^*$ the dual space to a Banach space $X$. The notation
$$
\div{v} = \nabla \cdot v = v_{i,i},  
%%= 
%%\sum_{i,j=1}^2 \frac{\partial v_i}{\partial x_i}, 
\ \ \ \ \ 
{\rm Div} \, \sigma = \nabla \cdot \sigma = 
\sigma_{ij,j} 
$$
stand for the divergence operators 
of the vector field $v \in L^2(\Omega)^2$ 
and 
of the tensor field $\sigma \in L^2(\Omega, \mathbb{S}^{d})$. 
An index that follows a comma indicates a derivative with respect to the corresponding component of the variable, 
and the summation convention over repeated indices is used.
For the scalar field $C \in H^1(\Omega)$, its gradient is denoted by
$\nabla C = (C_{,1}, C_{,2})$ and if $C\in H^2(\Omega)$ its conormal derivative is defined by 
$$\frac{\partial C}{\partial \nu} 
= \nabla C \cdot \nu.
$$

Recall, see~\cite[Theorem 2.15]{MOSBOOK}, 
that the embedding 
$i\colon E\to H^{1-\delta}(\Omega)^2$ is compact for 
$\delta \in (0,\frac{1}{2})$. 
By $\gamma_1\colon H^{1-\delta}(\Omega)^2\to L^2(\Gamma)^2$, 
we denote the trace operator, which is known to be continuous, see~\cite[Theorem 2.21]{MOSBOOK}.
Hence, the trace operator 
$\gamma=\gamma_1 i\colon E\to L^2(\Gamma)^2$ is compact. 
In what follows, the norm of $\gamma$ in $\mathcal{L}(E,L^2(\Gamma)^2)$ 
(the space of linear and bounded operators from $E$ 
into $L^2(\Gamma)^2$) is denoted by $\|\gamma\|$, 
and instead of $\gamma v$, we often write simply $v$.
We will also use the following special case 
of the Gagliardo--Nirenberg interpolation inequality, proof of which can be found 
in~\cite[Theorem 10.1]{FRIED}.
\begin{Lemma}\label{Lady}
	If $\Omega \subset \real^2$ 
	is a domain with $C^1$ boundary, then there exists
	a constant $M>0$ such that 
	\[	\|u\|_{L^4(\Omega)}\le M\|u\|_{L^2(\Omega)}^{1/2}
	\| u\|_{H^1(\Omega)}^{1/2} 
	\ \ \mbox{\rm for all} \ \ u \in H^1(\Omega). 	
	\]
\end{Lemma}

For a finite number $T>0$, we introduce 
the Bochner-Lebesque spaces 
\[
\mathbb{E} = \{\, v\in L^2(0,T;E) \mid v'\in L^2(0,T;E^*)\, \}
\]
and 
\[
\mathcal{W} 
= \{ \, C\in L^2(0,T;V) \mid C'\in L^2(0,T;V^*)\, \},
\]
where $v'$ and $C'$ denote the time derivatives in the sense of distributions.

We recall two useful results on evolution triples, proofs 
of which can by found in~\cite[Lemma~2.1]{TEMAN1} 
and~\cite[Corollary~4]{SIMON}, respectively.
\begin{Lemma}[Erhling]\label{Erhling}
	Let $X$, $Y$ and $Z$ be Banach spaces such that 
	$X$ is compactly embedded in $Y$, 
	and $Y$ is continuously embedded in $Z$. 
	Then, for every $\varepsilon >0$, there exists a constant $C(\varepsilon) > 0$ such that
	\begin{equation}\label{erh}
	\nonumber
	\|x\|_Y\leq \varepsilon \, \|x\|_X 
	+ C(\varepsilon) \, \|x\|_Z 
	\ \ \mbox{\rm for all} \ \ x \in X.
	\end{equation}
\end{Lemma}
\begin{Lemma}[Aubin-Lions] \label{AubinLions}
	Let $X$, $Y$ and $Z$ be reflexive Banach spaces and $X\subset Y \subset Z$ continuously with compact embedding $X\subset Y$,
	and $p$, $q\in (1,\infty)$. Then, for any $T>0$, the space 
	$$
	\{\, u \in L^p(0,T;X)\mid u'\in L^q(0,T;Z) \, \}
	$$ 
	is compactly embedded into $L^p(0,T;Y)$.
\end{Lemma}

In what follows, we denote by 
$\langle \cdot, \cdot \rangle_{X^*\times X}$ 
the duality pairing between a Banach space $X$ and its dual.

We recall the definitions of the generalized directional 
derivative and the generalized gradient of Clarke for a locally
Lipschitz function $\varphi \colon X \to \real$, where $X$ is a Banach space, see~\cite{CLARKE}. 
The generalized directional derivative of $\varphi$ at $x \in X$ in the
direction $v \in X$, denoted by $\varphi^{0}(x; v)$, is defined by
$$\displaystyle 
\varphi^{0}(x; v) =
\limsup_{y \to x, \ \lambda \downarrow 0}
\frac{\varphi(y+\lambda v) - \varphi(y)}{\lambda}.
$$
The generalized gradient of $\varphi$ at $x$, denoted by
$\partial \varphi(x)$, is a subset of a dual space $X^*$ given by
$$
\partial \varphi(x) = \{ \, 
\zeta \in X^* \mid \varphi^{0}(x; v) \ge 
{\langle \zeta, v \rangle}_{X^* \times X}  
\ \ \mbox{\rm for all} \ \ v \in X \, \}. 
$$

\smallskip

Finally, we recall the Green formula, proof of which can be found in e.g.~\cite[Theorem~2.25]{MOSBOOK}.
\begin{Lemma}\label{green2}
	Let $\Omega$ be an bounded domain in $\mathbb{R}^d$, $d= 2,3$ with Lipschitz boundary. 
	Then, the following formula holds
	\begin{equation*}
	\int_\Omega \sigma : \varepsilon (v)\,dx
	+ \int_\Omega {\rm Div}\,\sigma \cdot v\,dx
	= \int_\Gamma \sigma \nu \cdot v \, d\Gamma
	\end{equation*}
	for all $v\in H^1(\Omega)^d$ and $\sigma\in C^1(\bar{\Omega};\mathbb{S}^d)$, where $\varepsilon (u) = (\varepsilon_{ij}(u)) = (\frac{1}{2} (u_{i,j}+u_{j,i}))$,  $i$, $j = 1$, $2$, 
%	where $\mathbb{S}^d$ denotes the linear space 
%	of second order symmetric tensors on $\mathbb{R}^d$, $d = %2$, $3$ and $\sigma : \tau$ is a notation for product of %$\sigma$, 
%	$\tau\in \mathbb{S}^d$ given by $\sigma :  \tau=\sum_{i,j}^d %\sigma_{ij}\tau_{ij}$.
\end{Lemma}
Throughout the paper, we denote by $M$  a generic constant whose value may change from line to line.

\section{Classical formulation}\label{Statement}

In this section we provide the classical formulation 
of a model for miscible liquids which describe 
evolution of the velocity 
$u\colon \Omega \times (0,T)\to \real^2$, 
pressure $p \colon \Omega \times (0,T)\to \real$ 
and concentration 
$C \colon \Omega \times (0,T)\to \real$
of a viscous incompressible fluid
filling domain $\Omega$ with the time interval $(0, T)$. 

The model consists with
the incompressible Navier-Stokes equation 
modified by 
the (additional) Korteweg tensor. 
The classical stress tensor $\sigma$ 
for incompressible fluids is given 
by
\begin{equation}\label{stresstensor}
\sigma = -p\, I + 2 \, \nu_0 \, \varepsilon(u)
\ \ \mbox{in} \ \ \Omega \times (0,T), 
\end{equation}
where $I$ denotes the identity matrix and $\nu_0$ is the kinetic viscosity coefficient.
We suppose that the fluid is incompressible
\begin{equation}\label{eq_2}
\div{u} = 0 \ \ \mbox{in} \ \ \Omega \times (0,T),
\end{equation}
and governed by the Navier-Stokes equation for miscible fluids
\begin{equation}\label{eq_1}
\frac{\partial u}{\partial t} - \nu_0\, \Delta u 
+ (u\cdot \nabla ) u + \nabla p =  \Div K(C) +f
\ \ \mbox{in} \ \ \Omega \times (0,T),
\end{equation}
where $f\colon \Omega \times (0,T)\to \real^2$ 
denotes external forces field such as gravity and buoyancy, 
 and $K(C)=(K_{ij}(C))$ is the Korteweg stress tensor 
given by the following relations
\begin{equation}\label{T_def} 
K_{11} (C) = k \, \frac{\partial C}{\partial x_2}\frac{\partial C}{\partial x_2} \, ,  K_{22} (C) = k \, \frac{\partial C}{\partial x_1}\frac{\partial C}{\partial x_1} \, , K_{12}(C)=K_{21}(C)= -k \frac{\partial C}{\partial x_1} \frac{\partial C}{\partial x_2}
\end{equation}
where $k$ is a nonnegative constant.

We use a concentration function $C$
to represent and track the interface between liquids. 
The concentration function is transported by the velocity 
field $u$
\begin{equation}\label{eq_3}
\frac{\partial C}{\partial t} - d\, \Delta C 
+ u\cdot \nabla C = g \,C 
\ \ \mbox{in} \ \ \Omega \times (0,T),
\end{equation}
where $d >0$ is the coefficient of mass diffusion 
and $g$ represents the source term.
We assume also the homogeneous Neumann boundary conditon 
on the boundary $\Gamma$ for the concentration function
\begin{equation}
\frac{\partial C}{\partial \nu}= 0 
\ \ \mbox{on} \ \ \Gamma \times (0,T). \label{Neumann}
\end{equation} 

%The system of equation reads as follows.
%\begin{eqnarray}
%&&\label{eq_1}
%\frac{\partial u}{\partial t} - \nu_0\, \Delta u 
%+ (u\cdot \nabla ) u + \nabla p = \Div (T(C)) + f 
%\ \ \mbox{in} \ \ \Omega \times (0,T), \\[2mm]
%&&\label{eq_2}
%\div{u} = 0 \ \ \mbox{in} \ \ \Omega \times (0,T),\\[2mm]
%&&\label{eq_3}
%\frac{\partial C}{\partial t} - d\, \Delta C 
%+ u\cdot \nabla C = g \,C 
%\ \ \mbox{in} \ \ \Omega \times (0,T),\\[2mm]
%&&\label{eq_4}
%u(0)=u_0, \ C(0)=C_0 \ \ \mbox{in} \ \ \Omega.
%\end{eqnarray}

\noindent 
We supplement the system 
%%\eqref{eq_1}-\eqref{eq_4} 
with boundary and initial conditions. 
On the part $\Gamma_0$, 
we suppose adhesive boundary condition 
\begin{equation}\label{Gamma 0 con} 
u =0 \ \ \mbox{on} \ \ \Gamma_0\times (0,T).
\end{equation}
The following nonmonotone slip boundary condtion 
of frictional type with no leak 
is assumed on the part $\Gamma_1$
\begin{equation}\label{frictionlaw}
u_{\nu} =0, \ \ \
-\sigma_{\tau}\in \partial j(u_\tau) 
\ \ \mbox{on} \ \ \Gamma_1\times (0,T),
\end{equation}
where $\partial j$ denotes the generalized gradient 
of a prescibed locally Lipschitz function~$j$. 
%We assume that concentration does not affect 
%boundary conditions (\ref{Gamma 0 con}) 
%and (\ref{frictionlaw}). 
%
The boundary friction law (\ref{frictionlaw}) has been considered for the Navier-Stokes problems in~\cite{KalLuk,KASHI,OCHAL1,SZAFR}. 
Finally, the initial conditions for the velocity and concentration are prescibed
\begin{equation}\label{eq_4}
u(0)=u_0, \ C(0)=C_0 \ \ \mbox{in} \ \ \Omega.
\end{equation}

The classical formulation of the problem for miscible liquids is the following.

\begin{Problem}\label{p1}
Find
$u\colon \Omega \times (0,T)\to \real^2$, 
$p \colon \Omega \times (0,T)\to \real$ 
and 
$C \colon \Omega \times (0,T)\to \real$
such that (\ref{T_def})--(\ref{eq_4}) are satisfied.
\end{Problem}

In the next section we will study the weak formulation 
of Problem~\ref{p1}. 

We conclude this section with remarks on the Korteweg stress tensor which will be useful in next sections.
Using the notation
\begin{equation*}
\nabla C = \big(
\frac{\partial C}{\partial x_1}, 
\frac{\partial C}{\partial x_2}\big), \ \ \ 
\Delta C 
= \sum_{i=1}^2 \frac{\partial^2 C}{\partial x_i^2}
\end{equation*}
%We now look closer on the divergence of tensor $T(C)$, %appearing on the right-hand side of \eqref{eq_1}. Since $T(C)$ 
\noindent 
and formula~\eqref{T_def}, 
we calculate the first component of $\Div K(C)$ by 
\begin{eqnarray*}
&&
\frac{\partial{K_{11}}}{\partial x_1} +\frac{\partial{K_{12}}}{\partial x_2} = 2k\frac{\partial C}{\partial x_2} \frac{\partial^2 C}{\partial x_1 \partial x_2} -k \frac{\partial^2 C}{\partial x_1 \partial x_2} \frac{\partial C}{\partial x_2}-k\frac{\partial^2 C}{\partial x_2^2}\frac{\partial C}{\partial x_1}  \\[2mm]
&&
\quad 
=k(\frac{\partial C}{\partial x_2}\frac{\partial^2 C}{\partial x_1 \partial x_2}-\frac{\partial^2 C}{\partial x_2^2} \frac{\partial C}{\partial x_1} )\\[2mm]
&&
\qquad 
=k(\frac{\partial C}{\partial x_2} \frac{\partial^2 C}{\partial x_1\partial x_2} + \frac{\partial C}{\partial x_1}\frac{\partial^2 C}{\partial x_1^2}) -k\frac{\partial C}{\partial x_1} (\frac{\partial^2 C}{\partial x_1^2} + \frac{\partial^2 C}{\partial x_2^2}) \\[2mm]
&&
\qquad\quad
=\frac{k}{2}\frac{\partial}{\partial x_1}\|\nabla C\|^2 - k\frac{\partial C}{\partial x_1} \Delta C.
\end{eqnarray*}
Calculating, in the analogous way, the second component, 
we get
\[
\frac{\partial{K_{21}}}{\partial x_1} +\frac{\partial{K_{22}}}{\partial x_2}  =\frac{k}{2}\frac{\partial}{\partial x_2}\|\nabla C\|^2 - k\frac{\partial C}{\partial x_2} \Delta C.
\]
Hence, we have 
\begin{equation}\label{T_form}
\Div K(C) = 
\left(
\begin{matrix}
\displaystyle 
\frac{\partial{K_{11}}}{\partial x_1} +\frac{\partial{K_{12}}}{\partial x_2} \\ 
\displaystyle 
\frac{\partial{K_{21}}}{\partial x_1} +\frac{\partial{K_{22}}}{\partial x_2}
\end{matrix}
\right) =
\frac{k}{2} \nabla \|\nabla C\|^2 - k\, \Delta C\, \nabla C. 
\end{equation}
Using \eqref{T_form}, we easily obtain
\begin{eqnarray*}
&&\int_\Omega {\rm Div} K(C)\cdot v\, dx = \int_\Omega \Big(
\frac{k}{2} \nabla \|\nabla C\|^2 - k\, \Delta C\, \nabla C \Big)\cdot v\, dx \\[2mm]
&&
\quad =-\int_\Omega 
\frac{k}{2} \|\nabla C\|^2 \div v\, dx - \int_\Omega k\, \Delta C\, \nabla C\cdot v\, dx = -\langle  k\Delta C\nabla C,v\rangle_{L^2(\Omega)}
\end{eqnarray*}
for all $v\in E$ and $C\in V$.
Hence, we conclude 
\begin{equation}\langle \Div K(C),v\rangle_{E^*\times E}  = -\langle  k\Delta C\nabla C,v\rangle_{L^2(\Omega)}
\ \ \mbox{for all} \ \ v\in E, \, C\in V.
\end{equation}

\section{Variational formulation}\label{variational}

In this section we provide a variational formulation of Problem~\ref{p1} 
and state the main result of this paper.
%
%corresponding to the classical problem %\eqref{eq_1}--\eqref{frictionlaw} 
%and state the main theorem. 

We start by introducing the following 
%%bilinear and trilinear 
forms and formulating their properties.
The bilinear forms
$a_0\colon H^1(\Omega)^2\times H^1(\Omega)^2\to \mathbb{R}$,
$b_0\colon H^1(\Omega) \times H^1(\Omega) \to \mathbb{R}$ 
and
$c\colon H^1(\Omega)^2\times L^2(\Omega)\to \mathbb{R}$ 
are given by 
\begin{eqnarray*}
	&& a_0(u,v)=\frac{\nu_0}{2} \sum_{i,j=1}^2 \int_{\Omega} \Big(\frac{\partial u_i}{\partial x_j} 
	+ \frac{\partial u_j}{\partial x_i}\Big)
	\Big(\frac{\partial v_i}{\partial x_j} 
	+ \frac{\partial v_j}{\partial x_i}\Big) \, dx 
	\ \ \mbox{for} \ \ u,\, v\in H^1(\Omega)^2, \\[2mm]
	&& b_0(\xi,\eta) = d \int_{\Omega} \nabla \xi \cdot \nabla \eta\, dx \ \ \mbox{for} \ \ \xi,\, \eta \in H^1(\Omega), 
	\\[2mm]
	&& 
	c(v,q) = - \int_{\Omega} (\div v) \, q\, dx 
	\ \ \mbox{for} \ \, v\in H^1(\Omega)^2, 
	\, q\in L^2(\Omega).	
\end{eqnarray*}

\noindent 
We also define the trilinear forms 
$a_1\colon H^1(\Omega)^2\times H^1(\Omega)^2 \times H^1(\Omega)^2 \to \mathbb{R}$ and 
$b_1\colon H^1(\Omega)^2 \times H^1(\Omega) 
\times H^1(\Omega) \to \mathbb{R}$ by
\begin{eqnarray*}
&&
a_1(u,v,w)=\int_{\Omega} 
((u\cdot\nabla)\, v)\cdot w\, dx 
\ \ \mbox{for} \ \ u,\, v,\, w\in H^1(\Omega)^2, \\[2mm]
&& 
b_1(v,\xi,\eta) = \int_{\Omega} 
(v\cdot \nabla \xi) \, \eta \, dx \ \ \mbox{for} \ \ 
v\in H^1(\Omega)^2,\,  \xi,\, \eta \in H^1(\Omega).
\end{eqnarray*}

There exists $\alpha=\frac{1}{2}\nu_0 M_k>0$ , where $M_k$ is a constant arising from Korn inequality, such that
\begin{equation}\label{a_0b_0corecive}
a_0(u,u)\ge \alpha \, \|u\|_E^2, \quad 
b_0(\xi, \xi)= d\, \| \nabla\xi \|_{L^2(\Omega)}^2 
\end{equation}
for all $u \in H^1(\Omega)^2$ 
and $\xi \in H^1(\Omega)$.
Also, by the definition of space $E$, we have
\begin{equation}\label{c_zero}
c(v,q) = 0 \ \ \mbox{for} \ \ v\in E, 
\ q\in L^2(\Omega). 
\end{equation}
Moreover, we recall the properties of 
forms $a_1$ and $b_1$. They follow from Lemma~\ref{Lady} 
and Lemma~1.3(II) in~\cite{TEMAN1}.
\begin{Lemma} \label{lemma1}
{\rm (a)} For all $u$, $v$, $w \in E$, 
we have
\begin{eqnarray*}
&&
a_1(u,v,w)=-a_1(u,w,v), \\ [2mm]
&&
|a_1(u,v,w)|\leq M\|u\|_{L^2(\Omega)^2}^{1/2} \|u\|_E^{1/2} \|v\|_E \|w\|_{L^2(\Omega)^2}^{1/2} \|w\|_{E}^{1/2} \ \ \mbox{with} \ M>0, \\ [2mm]
&&
a_1(u,v,v)= 0.
\end{eqnarray*}
{\rm (b)} 
For all $u\in E$, $\eta$, $\zeta \in H^1(\Omega)$, we have
\begin{eqnarray*}
&&
b_1(u,\eta,\zeta)=-b_1(u,\zeta,\eta),\\ [2mm]
&&
|b_1(u,\eta,\zeta)|\le M\|u\|_{L^2(\Omega)^2}^{1/2} \|u\|_E^{1/2} \|\eta\|_{ H^1(\Omega)} \|\zeta\|_{L^2(\Omega)}^{1/2} \|\zeta\|_{ H^1(\Omega)}^{1/2}\ \ \mbox{with} \ M>0, \\ [2mm]
&&
b_1(u,\eta,\eta)= 0.
\end{eqnarray*}
\end{Lemma}

\noindent 

Furthermore, we introduce operators 
$A_0$, $A_1\colon E\to E^*$, 
$B_0\colon V\to V^*$ and 
$B_1 \colon E\times V\to V^*$ defined by 
\begin{eqnarray}
&&\nonumber
\label{operators} \langle A_0 u, 
v\rangle_{E^*\times E}=a_0(u,v), \ \ \  
\langle A_1 u,v\rangle_{E^*\times E}=a_1(u,u,v) 
\ \ \mbox{for} \ \ u,\, v\in E, \\[2mm]
&&\nonumber 
\langle B_0 C,\eta\rangle_{V^*\times V}=b_0(C,\eta), \ \ \  \langle B_1(u,C),\eta\rangle_{V^*\times V}=b_1(u,C,\eta) 
\ \ \mbox{for} \ \ u \in E, \, C, \, \eta \in V.
\end{eqnarray}

Assume now that $u$, $p$ and $C$ are sufficiently smooth 
functions which solve Problem~\ref{p1}. 
Let $v \in E$ be sufficiently smooth and $t \in (0, T)$.
Using the Green formula 
of Lemma~\ref{green2}, combined with the definition of \eqref{stresstensor}, 
similarly as in~\cite{SZAFR}, we obtain 
the following equality
\begin{eqnarray}
&&\nonumber\int_{\Omega} p \, \div v \, dx 
+ \nu_0 \int_\Omega \varepsilon(u)\colon \varepsilon(v)\, dx  + \int_\Omega (-\nabla p + \nu_0 \Delta u) \cdot v \, dx 
= \int_{\Gamma} \sigma \nu \cdot v \, dx. \label{after_green} \nonumber
\end{eqnarray}
%%for all $v\in E$. 
From the definition of forms $a_0$ and $c$, 
we have
\begin{equation}
\int_\Omega (\nabla p - \alpha \Delta u) 
\cdot v \, dx  
= a_0(u,v) + c(v,p) -  \int_{\Gamma} \sigma \nu \cdot v\, d\Gamma. \label{dziwna_formula}
\end{equation}
%%for all $v\in E$. 
\noindent 
We now multiply \eqref{eq_1} by $v\in E$. Exploiting definitions of operators 
$A_0$, $A_1$, using \eqref{c_zero} and \eqref{dziwna_formula}
we deduce 
\begin{eqnarray}
&&\nonumber
 \langle u'(t) + A_0 u(t) + A_1 u(t),v\rangle_{E^*\times E} - \int_{\Gamma} \sigma \nu \cdot v\, d\Gamma  \\[2mm]
&& 
\quad = \langle \Div K(C), v\rangle_{E^*\times E} 
+ \langle f(t), v\rangle_{E^*\times E} . \label{var_1}
\end{eqnarray}
%%for all $v\in E$.
\noindent
Next, we use the orthogonality relation 
$\sigma\nu\cdot v 
= \sigma_\nu v_\nu + \sigma_\tau\cdot v_\tau 
= \sigma_\tau\cdot v_\tau$ 
%%for all $v\in E$, 
and \eqref{frictionlaw} to arrive at the equality
\begin{equation}
\label{var2} \langle u'(t) + A_0 u(t) + A_1 u(t),v\rangle_{E^*\times E} +\langle \xi(t),v\rangle_{L^2(\Gamma_1)^2} =  \langle \Div{K(C(t))} + f(t),v\rangle_{E^*\times E},
\end{equation}
where $\xi \in L^2(0,T;L^2(\Gamma_1)^2)$, 
$\xi(t) \in \partial j(u_\tau(t))$ 
for a.e. $t \in (0, T)$.
On the other hand, we multiply \eqref{eq_3} 
by $\eta\in V$, using \eqref{Neumann} we find
\begin{equation}
\langle C'(t)+B_0C(t)+B_1(u(t),C(t)),\eta\rangle_{V^*\times V} = \langle g\, C(t),\eta\rangle_{L^2(\Omega)}. \label{var3}
\end{equation}

Summarizing, we obtain the following system 
of equations and inclusion which is the variational formulation of Problem~\ref{p1}.
\begin{Problem}\label{PVar}
	Find $u\in \mathbb{E}$ and $C\in \mathcal{W}$ such that 
	there exists $\xi \in L^2(0, T; L^2(\Gamma_1)^2)$ and
	\begin{eqnarray*}
	&& \nonumber
	\langle u'(t) + A_0 u(t) + A_1 u(t),v\rangle_{E^*\times E} +\langle \xi(t),v\rangle_{L^2(\Gamma_1)^2}\\[2mm]
	&&\nonumber 
	\qquad \ \ 
	=  \langle \Div{K(C(t))} + f(t),v\rangle_{E^*\times E} 
	\ \ \mbox{\rm for all} \ \ v\in E, 
	\ \mbox{\rm a.e.} \ t\in (0,T),\\[2mm]
	&&
	\xi(t) \in \partial j(u_\tau(t))
	\ \ \mbox{\rm for a.e.} \ t\in (0,T), \\ [2mm]
	&&\nonumber
	\langle C'(t)+B_0C(t)+B_1(u(t),C(t)), 
	 \eta\rangle_{V^*\times V} = \langle g\, C(t),\eta\rangle_{L^2(\Omega)} \\[2mm]
	&&\nonumber
	\qquad \ \ 
	\mbox{\rm for all} \ \ \eta\in V, \ \mbox{\rm a.e.} 
	\ t\in (0,T),\\[2mm]
	&&\nonumber
	u(0)=u_0, \ C(0)=C_0.
	\end{eqnarray*}
\end{Problem}

We need the following hypotheses.

\medskip

\noindent
$\underline{H(j)}:$ 
\ $j\colon \Gamma_1 \times \real^2 \to \real$ is such that 

\smallskip

{\rm (a)} \
$j(\cdot,\xi)$ is measurable for all $\xi \in \real^2$, $j(\cdot,0)\in L^2(\Gamma_1)$,

\smallskip

{\rm (b)} \
$j(x,\cdot)$ is locally Lipschitz for a.e. $x\in \Gamma_1$,

\smallskip

{\rm (c)} \ 
$\eta \cdot s\ge 0$ for all 
$\eta\in \partial j(x, s)$, $s\in \real^2$, 
a.e. $x \in \Gamma_1$,

\smallskip

{\rm (d)} \
$\|\zeta\|\le m_0\, (1+\|\xi\|)$ for all 
$\xi \in \real^2$, $\zeta \in\partial j(x,\xi)$, 
a.e $x\in\Gamma_1$ with $m_0>0$,

\smallskip

{\rm (e)} \ 
$(\zeta_1 - \zeta_2) \cdot (\xi_1 - \xi_2) \ge -
m_1 \| \xi_1 - \xi_2 \|^2$ 
for all $\zeta_i \in \partial j (x, \xi_i)$, 
$\xi_i \in \mathbb{R}^2$, 

\smallskip

\qquad $i=1$, $2$, 
a.e. $x \in \Gamma_1$ with $m_1 \ge 0$.

\medskip

\noindent 
$\underline{(H_0)}$: \ $d$, $k$, $\nu_0>0$, 
$g\in L^\infty(\Omega)$, $g\ge 0$, 
$f\in L^2(0,T;E^*)$, $u_0\in E$, $C_0\in V$.

\medskip 

Our main result of this paper on a unique solvability of 
Problem~\ref{PVar} reads as follows.
\begin{Theorem}\label{maintheorem}
Under hypotheses $H(j)$$(a)$\mbox{--}$(d)$ and $(H_0)$, Problem~\ref{PVar} has a solution such that 
$C\in L^\infty(0,T;H^1(\Omega))$. If, in addition, 
$H(j)$$(e)$ holds, then the solution to Problem~\ref{PVar} 
is unique.
\end{Theorem}

\section{Proof of the main result}\label{proof}

In this section we provide the proof 
of Theorem~\ref{maintheorem}.
For the existence, we use the regularized Galerkin method. 
To this end, we define the regularization of the mulitivalued term as follows.

Let $\rho\in C_0^{\infty}(\mathbb{R}^2)$ be the mollifier such that 
$\rho \ge 0$ on $\mathbb{R}^2$, 
$\mbox{supp}\,\rho \subset [-1,1]^2$ and $\int_{\mathbb{R}^2} \rho \, dx =1$. 
We define $\rho_m(x) = m^2 \rho(mx)$ for $m \in \mathbb{N}$.
Then $\mbox{supp}\, \rho_m \subset [-\frac{1}{m}, \frac{1}{m}]^2$ 
for all $m \in \mathbb{N}$. 
Consider functions 
$j_{m} \colon \Gamma_1\times\mathbb{R}^2 \to \mathbb{R}$ 
defined by
$$
j_{ m}(x,\xi) = \int_{supp\,\rho_m} \rho_m(z) j (x,\xi-z)\,dz 
\ \ \mbox{for} \ \ (x,\xi) \in\Gamma_1\times\mathbb{R}^2.
$$
We observe that since
$j_{ m}(x,\cdot)\in C^\infty (\mathbb{R}^2)$ for all $x\in \Gamma_1$, therefore $\partial j_{m}(x,\xi)$ 
reduces to a single element. 
We write 
$\partial j_{ m}(x,\xi(t)) = \{ D_u j_{ m}(x,\xi(t))\}$ for all $\xi(t)\in E$, where $D_u j_m$ represents the derivative of $j_m(x,\cdot)$. Moreover, it is easy to see that $j_m$ satisfies the growth condition $H(j)(d)$.

Using the separability of the space $E$, 
we may write a basis of $E$ as $\{\varphi_1,\varphi_2,\ldots\}$.
We choose in $V$ 
a special basis $\{\psi_1,\psi_2,\ldots\}$ 
of eigenvectors of the $-\Delta$ eigenvalue problem associated with zero Neumann boundary condition,
see \cite[Theorem 6.1.31]{Gasinski}.

We define finite dimensional subspaces 
$E^m={\rm span} \{\varphi_1,\ldots,\varphi_m\}$ of $E$, and $V^m={\rm span} \{\psi_1,\ldots,\psi_m\}$ of $V$ 
for $m\ge 1$. 
Let $u_{0m}$, $C_{0m}$ be such that 
 $u_{0m}\to u_0$ in $H$ 
and 
$C_{0m}\to C_0$ in $L^2(\Omega)$ 
with $u_{0m}\in E^m$ and $C_{0m}\in V^m$ 
for $m\ge 1$.
Next, for a fixed $m\ge 1$, consider 
the following problem in finite dimensional spaces.
\begin{Problem}\label{Gal1}
	Find $u_m\in L^2(0,T;E^m)$ with $u_m'\in L^2(0,T;E^m)$ 
	and $C_m\in L^2(0,T;V^m)$ with $C_m'\in L^2(0,T;V^m)$ 
	such that
	\begin{eqnarray}
	&&\nonumber
	\langle u_m'(t) + A_0u_m(t) + A_1u_m(t),v_m\rangle_{E^{m*}\times E^m} + \langle D_uj_m(u_{m\tau}(t)),v_m\rangle_{L^2(\Gamma_1)^2}  \\[2mm]
	&&\label{galeq1}
	\quad 
	=\langle f(t) + \Div K(C_m(t)),v_m\rangle_{E^{m*}\times E^m} \ \mbox{\rm for all} \ v_m\in E^m, 
	\ \mbox{\rm a.e.} \ t\in (0,T),\\[2mm]
	&&\nonumber
	\langle C_m'(t) + B_0C_m(t)+B_1(u_m(t),C_m(t)), \eta_m\rangle_{V^{m*}\times V^m} = \langle g\, C_m(t),\eta_m\rangle_{L^2(\Omega)} \\[2mm]
	&&\label{galeq2}
	\quad \mbox{\rm for all} \ \eta_m\in V^m, 
	\ \mbox{\rm a.e.} \ t\in (0,T),\\[2mm]
	&&\nonumber
	u_m(0)=u_{0m}, \ C_m(0)=C_{0m}.
	\end{eqnarray}
\end{Problem}

\noindent
We introduce the new variable 
$z_m(t)=(u_m(t), C_m(t))$ 
and the space $Y^m=E^m\times V^m$, 
and rewrite 
Probem~\ref{Gal1} as follows: 
find $z_m\in L^2(0,T;Y^m)$ with $z_m'\in L^2(0,T;Y^m)$ 
such that $z_m(0) = (u_{0m}, C_{0m})$ and
\begin{equation}\label{gal_z}
\langle z_m'(t),y_m\rangle = 
\langle Az_m(t),y_m\rangle 
\ \ \mbox{for all} \ \ y_m\in Y^m, \ \mbox{a.e.} \ t\in (0,T),
\end{equation} 
where $y_m = (v_m, \eta_m)$ and
\begin{eqnarray*} 
&&
\hspace{-0.5cm}
\langle Az_m (t), y_m\rangle = \\ [2mm]
&&
 = \begin{pmatrix} \langle -A_0u_m(t) - A_1u_m(t) +\Div K(C_m(t)) +f(t),v_m\rangle - \langle D_uj_m(u_{m\tau}(t)),v_m\rangle_{L^2} \\
  \langle -B_0C_m(t)-B_1(u_m(t),C_m(t)),\eta_m\rangle + \langle g\, C_m(t),\eta_m\rangle
 \end{pmatrix} .
\end{eqnarray*}

Solvability of the problem~\eqref{gal_z} on a small time
interval $(0,T_1)$ follows from the Carath\'eodory existence theorem. We now show a priori estimates to extend 
the solution on the whole interval $(0,T)$. 
First, we test equation~\eqref{galeq2} with 
$\eta_m=C_m(t)$ and observe that from Lemma~\ref{lemma1}(b), 
we have 
$b_0(u_m(t),C_m(t),C_m(t))=0$ for a.e. $t \in (0, T)$.
Hence, we obtain
\begin{eqnarray}
\label{est1} 
\frac{1}{2}\frac{d}{dt}\|C_m(t)\|_{L^2(\Omega)}^2 
+ d\, \|\nabla C_m(t)\|_{L^2(\Omega)}^2 \le \|g\|_{L^\infty(\Omega)}\|C_m(t)\|_{L^2(\Omega)}^2.
\end{eqnarray}
Integrating~\eqref{est1} over $(0,t)$ for $t\in (0,T)$, 
we get
\begin{equation}\label{est3}
\frac{1}{2}\|C_m(t)\|_{L^2(\Omega)}^2 
+ d\int_0^t \|\nabla C_m(s)\|_{L^2(\Omega)}^2 \, ds \le \|g\|_{L^\infty(\Omega)} \int_0^t\|C_m(s)\|_{L^2(\Omega)}^2 \,ds + M. 
\end{equation}
Using the Gronwall lemma, from the last inequality, 
we deduce that 
%%\eqref{est3} that
\begin{equation}\label{est4}
\|C_m\|_{L^\infty(0,T;L^2(\Omega))} \le M
\end{equation}
and putting \eqref{est4} in \eqref{est3},
we obtain
\begin{equation}\label{bound_CinftyH1}
\|C_m\|_{L^2(0,T;H^1(\Omega))} \le M.
\end{equation}

\noindent
Now, we take $v_m=u_m(t)$ in equality \eqref{galeq1}. 
Using coercivity of $A_0$ stated in \eqref{a_0b_0corecive}, condition $H(j)(d)$ 
and Lemma~\ref{lemma1}(a), we find 
\begin{eqnarray}
&&\nonumber
\frac{1}{2}\frac{d}{dt}\|u_m(t)\|_H^2 
+ \alpha \, \|u_m(t)\|_E^2 \le m_0\, (1+\|u_m(t)\|_{L^2(\Gamma_1)^2})
\|u_m(t)\|_{L^2(\Gamma_1)^2} \\[2mm]
&&\quad 
+ \langle \Div K(C_m(t)), u_m(t)\rangle + \|f(t)\|_{E^*}\|u_m(t)\|_E \label{est5}
\end{eqnarray}
for a.e. $t \in (0, T)$.
Using \eqref{T_form} and the Cauchy inequality with 
$\varepsilon > 0$ in \eqref{est5}, we obtain  
\begin{eqnarray}
&&\nonumber
\frac{1}{2}\frac{d}{dt}\|u_m(t)\|_H^2 
+ \alpha \, \|u_m(t)\|_E^2 
\le M \, \|u_m(t)\|_{L^2(\Gamma_1)}^2 \\[2mm]
&& \quad + \, \langle \Delta C_m(t)\nabla C_m(t), u_m(t)\rangle_{L^2(\Omega)^2} 
+ \varepsilon \, \|u_m(t)\|_V^2 +M \label{est6}
\end{eqnarray}
for a.e. $t \in (0, T)$.
From Lemma~\ref{Erhling}, there exists $M(\varepsilon) > 0$ 
such that 
$$
\|u_m(t)\|_{L^2(\Gamma_1)}^2 
\le 2\, \varepsilon \, \|u_m\|_E^2 + M(\varepsilon)\,\|u_m(t)\|_H^2.
$$ 
Using this inequality in \eqref{est6}, we have
\begin{eqnarray}
&&\nonumber\label{est7}
\frac{1}{2}\frac{d}{dt}\|u_m(t)\|_H^2 + (\alpha-3\varepsilon) \|u_m(t)\|_E^2  \\[2mm]
&&\quad  
\le \langle k\, \Delta C_m(t)\, \nabla C_m(t),u_m(t)\rangle_H + M(\varepsilon) \|u_m(t)\|_H^2 + M.
\end{eqnarray}

Subsequently, we take 
$\eta_m=-k\,\Delta C_m(t)$ in \eqref{galeq2} to find
\begin{eqnarray}
&&\nonumber
\frac{1}{2}\frac{d}{dt} \|\nabla C_m(t)\|_{L^2(\Omega)^2}^2 
+ d\, k \, \|\Delta C_m(t)\|_{L^2(\Omega)}^2 \\[2mm]
&&\label{est8}
\quad - k \, \langle \Delta C_m(t),\nabla C_m(t)u_m(t)\rangle_{L^2(\Omega)} 
\le k\, \|g\|\|\nabla C_m(t)\|_{L^2(\Omega)^2}^2.
\end{eqnarray}
Next, adding \eqref{est7} and \eqref{est8}, we get 
\begin{eqnarray}
&&\nonumber
\frac{1}{2}\frac{d}{dt} \left( \|u_m(t)\|_H^2 + \|\nabla C_m(t)\|_{L^2(\Omega)^2}^2 \right) + (\alpha -3 \varepsilon)\|u_m(t)\|_E^2  \\[2mm]
&&\label{est9}
\quad 
+\|\Delta C_m(t)\|_{L^2(\Omega)^2}^2 
\le M\, \|\nabla C_m(t)\|_{L^2(\Omega)^2}^2 
+ M(\varepsilon)\, \|u_m(t)\|_H^2 + M.
\end{eqnarray}
Choosing $\varepsilon > 0$ sufficiently small, 
integrating \eqref{est9} over $(0,t)$ for $t\in (0,T)$, 
from the Gronwall lemma, we deduce
\begin{eqnarray}
&& \label{bound_uinf}\|u_m\|_{L^\infty(0,T;H)}\le M, 
\\[1mm]
&& 
\label{bound_u2}\|u_m\|_{L^2(0,T;E)}\le M,\\[1mm]
&& \label{bound_CinfH}\|C_m\|_{L^\infty(0,T;H^1(\Omega))}
\le M, \\[1mm]
&& \label{bound_CH2}\|C_m\|_{L^2(0,T;H^2(\Omega))}\le M,
\end{eqnarray}
where \eqref{bound_CH2} holds due to \cite[Theorem 3.1.2.3]{Grisvald}.
Now, we estimate $\|C_m'\|_{L^2(0,T;L^2(\Omega))}$. 
To this end, using Lemmata~\ref{lemma1}(b) 
and~\ref{Lady}, we take $\eta_m=C_m'(t)$ in \eqref{galeq2} to find
\begin{eqnarray}
&&\nonumber
\hspace{-1.0cm}
\|C_m'(t)\|_{L^2(\Omega)}^2 \le 
d\, \|\Delta C_m(t)\|_{L^2(\Omega)}\|C_m'(t)\|_{L^2(\Omega)}  \\[2mm]
&&\nonumber
\hspace{-0.5cm} 
+\|u_m(t)\|_{L^4(\Omega)}\|\nabla C_m(t)\|_{L^4(\Omega)} 
\|C_m'(t)\|_{L^2(\Omega)} + \|g(t)\|_{L^\infty(\Omega)} \|C_m(t)\|_{L^2(\Omega)} \|C_m'(t)\|_{L^2(\Omega)} \\[2mm]
&& \nonumber
\le M\, \|C_m(t)\|_{H^2(\Omega)}\|C_m'(t)\|_{L^2(\Omega)} +\\[2mm]
&&\label{est10}
 +M\|u_m(t)\|_H^{1/2}\|u_m(t)\|_E^{1/2} \|C_m(t)\|_{H^1(\Omega)}^{1/2}\|\nabla C_m(t)\|_{H^1(\Omega)}^{1/2} \|C_m'(t)\|_{L^2(\Omega)}
\end{eqnarray}
for a.e. $t \in (0, T)$.
From \eqref{est10}, we infer that
\begin{equation}
\label{bound_C'}\|C_m'\|_{L^2(0,T;L^2(\Omega))}\le M.
\end{equation}

Next, we estimate the term $\Div K(C_m)$.
We observe that 
\begin{equation}
\Div K(C) = \sum_{i,j,k=1}^2 a_{ijk} D_i\, (D_j C \,D_k C), 
\label{T_sum}
\end{equation}
where $a_{ijk}$ are constants for $i$, $j$, 
$k=1$, $2$ and 
$D_l=\frac{\partial}{\partial x_l}$ 
for $l=1$, $2$.
We estimate one term in (\ref{T_sum}) and find
%
% one of the term in the sum on the right hand 
%side of \eqref{T_sum}
\begin{eqnarray}
&&\nonumber
\hspace{-1.0cm}
\| D_i(D_jC_m(t) D_k C_m(t)\|_{E^*} = 
\sup_{\|v\|_E=1} 
|\langle D_i(D_jC_m(t) D_k C_m(t),v\rangle_{E^*\times E}| 
\\[2mm]
&&
\hspace{-0.5cm}
\nonumber
\le \|D_jC_m(t)D_kC_m(t)\|_{L^2(\Omega)} \le  \|D_jC_m(t)\|_{L^4(\Omega)} \|D_k C_m(t)\|_{L^4(\Omega)}   \\[3mm]
&&\label{est11}
\le M\, \|D_jC_m(t)\|_{L^2(\Omega)}^{1/2}
\|D_jC_m(t)\|_{H^1(\Omega)}^{1/2} \|D_kC_m(t)\|_{L^2(\Omega)}^{1/2}\|D_kC_m(t)\|_{H^1(\Omega)} ^{1/2}.
\end{eqnarray}
From bounds~\eqref{bound_uinf}--\eqref{bound_CH2}, 
\eqref{T_sum} and \eqref{est11}, we have 
\begin{equation}
\|\Div K(C_m)\|_{L^2(0,T;E^*)} \le M. \label{bound_T}
\end{equation}
%From \eqref{bound_uinf}, \eqref{bound_u2}, \eqref{bound_T},  and H(j)(c) we see that 
%\begin{equation}
%\|u_m'\|_{L^2(0,T;E^*)} \le M.
%\end{equation}

Furthermore, from~\eqref{bound_CinftyH1}, \eqref{bound_uinf}--\eqref{bound_CH2} and \eqref{bound_C'}, 
we find elements 
$u \in L^\infty(0,T;H) \cap L^2(0,T;E)$ 
and $C \in L^\infty(0,T;H^1(\Omega))
\cap L^2(0,T;H^2(\Omega))$ 
such that, up to a subsequence, we get
\begin{eqnarray}
&&
u_m \to u \ \ \mbox{weakly}^* \ \ \mbox{in} \ L^\infty(0,T;H), \label{u_mtouinfty}
\\[1mm]
&&
u_m\to u \ \ \mbox{weakly} \ \ \mbox{in}  \ L^2(0,T;E),\label{u_mtouL2}
\\[1mm]
&&
C_m\to C \ \ \mbox{weakly}^* \ \ \mbox{in} \ L^\infty(0,T;H^1(\Omega)),\label{C_mtoCinfty}
\\[1mm]
&&
C_m \to C \ \ \mbox{weakly} \ \ \mbox{in} \ L^2(0,T;V), \label{C_mtoCL^2}
\end{eqnarray}
as $m \to \infty$.
By the definition of operator $A_1$ 
and~\cite[Lemma~3.4]{TEMAN1}, 
we have 
\begin{equation} \label{2d}
\|A_1(u_m)\|_{L^2(0,T;E^*)} \le M\|u_m\|_{L^\infty(0,T;H)} \|u_m\|_{L^2(0,T;E)}.
\end{equation}

\noindent
From \eqref{bound_uinf}, \eqref{bound_u2},  \eqref{bound_T},
\eqref{2d} and the definition 
of operator $A_0$, we infer that 
\begin{equation}\label{bound_u'}
\|u_m'\|_{L^2(0,T;E^*)}\le M,
\end{equation}
and hence 
\begin{equation} 
u'_m \to u'\ \ \mbox{weakly in} \ \ L^2(0,T;E^*), 
\ \ \mbox{as} \ \ m \to \infty . \label{u'mtou}
\end{equation}

\noindent
By Lemma~\ref{AubinLions}, we know that the embedding $\mathbb{E} \subset L^2(0,T;H)$ is compact, so from \eqref{bound_u2} and \eqref{bound_u'}, we have 
\begin{equation}\label{u_strong}
u_m \to u \ \ \mbox{in} \ \ L^2(0,T;H), 
\ \ \mbox{as} \ \ m \to \infty.
\end{equation}

\noindent 
Since the operator $A_0\colon E\to E^*$ 
is linear and continuous, 
so is its Nemytskii operator which is denoted in the same way.
Therefore, we find that 
\begin{equation}
A_0 u_m \to A_0 u \ \mbox{\rm weakly in} \ L^2(0,T;E^*),  \ \ \mbox{as} \ \ m \to \infty.\label{A_0conv}
\end{equation}
\noindent
From \eqref{bound_uinf} and \eqref{bound_u2} 
and the technique used in~\cite[Lemma III.3.2]{TEMAN1}, 
we have
\begin{equation}
A_1(u_m)\to A_1(u) \ \ \mbox{weakly in} \ \ L^2(0,T;E^*),
\ \ \mbox{as} \ \ m \to \infty. \label{A_1conv}
\end{equation}

We use the fact that the embedding 
$\mathcal{W} \subset L^2(0,T;H^1(\Omega))$
is compact. 
From \eqref{bound_C'}, \eqref{bound_CinftyH1} and Lemma~\ref{AubinLions}, 
we deduce
\begin{equation}\label{C_strong}
C_m \to C \ \ \mbox{in} \ \ L^2(0,T;H^1(\Omega)), 
\ \ \mbox{as} \ \ m \to \infty.
\end{equation}
\noindent
Moreover, from \eqref{bound_T} and \eqref{C_strong}, 
we see that
\begin{equation}
\Div K(C_m) \to \Div K(C) \ \ \mbox{weakly in} \ \ L^2(0,T;E^*),
 \ \ \mbox{as} \ \ m \to \infty. \label{DivCconv}
\end{equation}

Next, by the compactness of the trace operator 
from $\mathbb{E}$ to $L^2(0,T;L^2(\Gamma_1)^2)$, 
it follows
\[
u_{m\tau} \to u_\tau \ \ \mbox{in} \ \ L^2(0,T;L^2(\Gamma_1)^2), 
\ \ \mbox{as} \ \ m \to \infty.
\]

\noindent
Hence, by passing to a next subsequence, if necessary, 
we have
\begin{equation} \label{1}
u_{m\tau}(t) \to u_\tau(t) 
\ \ \mbox{in} \ \ L^2(\Gamma_1)^2 
\ \ \mbox{for a.e.} \ t\in (0,T),
\ \ \mbox{as} \ \ m \to \infty.
\end{equation}

On the other hand, by hypothesis $H(j)$(d) 
and~\eqref{bound_u2}, 
we may suppose that
\begin{equation} \label{2}
D j_{ m} (u_{m\tau}(\cdot)) \to \xi \ \ \mbox{weakly in} \ \ L^2(0,T;L^2(\Gamma_1)^2),
\ \ \mbox{as} \ \ m \to \infty
\end{equation}
with $\xi \in L^2(0,T;L^2(\Gamma_1)^2)$.
Now, we are in a position to use convergences (\ref{1}) and (\ref{2}), and apply the Aubin-Cellina convergence theorem, 
see~\cite[Theorem~1, p.60]{AUBIN} to the inclusion 
$$
D j_{ m}(u_{m\tau}(t)) \in \partial j_{m}(u_{m\tau}(t)) \ \ \mbox{\rm for a.e.} \ \ t\in (0,T).
$$ 
%{\color{red}\bf 
%[Od poczatku pod znakiem $\partial j_m$ powinno %byc 
%$u_{m\tau}(t)$?]}{\color{blue} [ $u_\tau =u - %u_\nu \nu = u$, ale i tak poprawione]}
%
We deduce that 
$$
\xi(t) \in {\rm \bar{co}} \, \partial j (u_\tau(t)) 
= \partial j(u_\tau(t)) 
\ \ \mbox{\rm for a.e.} \ \ t\in (0,T),
$$
where ${\rm {\bar{co}}}$ denotes the closure of the convex hull of a set.  
The last equality follows from the fact that the values of the generalized subgradient are closed and convex sets, 
see~\cite[Proposition~3.23]{MOSBOOK}. 

In a similar way, as in \eqref{A_0conv}, 
by linearity and continuity of operator 
$B_0$, by using \eqref{C_strong}, 
we have
\begin{equation}
B_0 C_m \to B_0 C \ \ \mbox{\rm weakly in} \ \ L^2(0,T;V^*), 
\ \ \mbox{as} \ \ m \to \infty. \label{B_0conv}
\end{equation}

\noindent
Also, from \eqref{u_strong} and \eqref{C_strong}, we obtain
\begin{equation}
B_1(u_m,C_m)\to B_1(u,C) \ \ \mbox{weakly in} \ \ L^2(0,T;V^*),
\ \ \mbox{as} \ \ m \to \infty. \label{B_1conv}
\end{equation}

From \eqref{bound_C'} we infer that 
\begin{equation}
C_m' \to C \ \ \mbox{weakly in} \ \ L^2(0,T;L^2(\Omega)). \label{C'conv}
\end{equation}

\noindent 
Thus, using convergences \eqref{u'mtou}, \eqref{A_0conv}, \eqref{A_1conv}, \eqref{DivCconv} and \eqref{2}, 
we pass to the limit in \eqref{galeq1} 
and
using standard techniques, see \cite[p.739]{OCHAL1} we obtain
\begin{eqnarray*}
&&
\langle u'(t) + A_0 u(t) + A_1 u(t), v\rangle_{E^*\times E} + \langle \xi(t),v\rangle_{L^2(\Gamma_1)^2}\\[2mm]
&&\nonumber 
\qquad \ \ 
=  \langle \Div{K(C(t))} + f(t), v\rangle_{E^*\times E} 
\ \ \mbox{\rm for all} \ \ v\in E, 
\ \mbox{\rm a.e.} \ t\in (0,T).
\end{eqnarray*}
Moreover, using \eqref{B_0conv}--\eqref{C'conv}, 
we pass to limit in \eqref{galeq2} and get 
\begin{eqnarray*}
&&\nonumber
\langle C'(t)+B_0C(t)+B_1(u(t),C(t)), 
\eta\rangle_{V^*\times V} = \langle g\, C(t),\eta\rangle_{L^2(\Omega)} \\[2mm]
&&\nonumber
\qquad \ \ 
\mbox{\rm for all} \ \ \eta\in V, \ \mbox{\rm a.e.} 
\ t\in (0,T).
\end{eqnarray*}
Since the mapping $\mathbb{E} \ni w  \to w(0) \in H$ is linear and continuous, from \eqref{u_mtouL2} and \eqref{u'mtou}, we have $u_m(0) \to
u(0)$ weakly in $H$, which together 
with $u_{0m} \to u_0$ in $H$ entails $u(0) = u_0$. 
Similarily, 
since $\mathcal{W}\ni \zeta \to \zeta(0)\in L^2(\Omega)$ is linear and continuous, we obtain $C(0)=C_0$. 
Finally, taking into account that 
$\xi(t) \in \partial j(u_\tau(t))$ 
for a.e. $t\in (0,T)$, we conclude 
that $u \in \mathbb{E}$ and 
$C \in \mathcal{W}$ is a solution to Problem~\ref{PVar}. Observe, that by \eqref{C_mtoCinfty}, we have the additional regularity $C\in L^\infty(0,T;H^1(\Omega))$. 
This concludes the existence proof.
%%$\hfill{ \square}$

\smallskip

We pass to the proof of uniqueness of solution to
Problem~\ref{PVar}. To show uniqueness of solution, 
we assume additionally the regularity of function $j$ stated in $H(j)(e)$.

%In this section we omit for the clarity of presentation the %time dependence notation of all terms.
Let $(u_1, C_1)$ and $(u_2, C_2)$ be two solutions 
of Problem~\ref{PVar}. Set $u=u_1-u_2$ and $C=C_1-C_2$. 
Using property~\eqref{T_form}, we obtain that $(u,C)$ is a solution to the following problem.
\begin{eqnarray}
&& \nonumber
\hspace{-1.0cm}
\langle u'(t) + A_0 u(t) + A_1 u_1(t)-A_1 u_2(t),v\rangle +\langle \xi^1(t)-\xi^2(t),v\rangle\\[2mm]
&&\label{uni1} 
\qquad \hspace{-1.0cm}
=  -k\langle \Delta C_1(t)\nabla C_1(t)-\Delta C_2(t)\nabla C_2(t),v\rangle  \ \ \mbox{for all} \ v\in E, \ \mbox{a.e.} \ t\in (0,T),\\[2mm]
&&\nonumber
\hspace{-1.0cm}
\langle C'(t)+B_0C(t)+B_1(u_1(t),C_1(t)-B_1(u_2(t),C_2(t),\eta \rangle= \langle g\, C(t),\eta \rangle \\[2mm]
&&\label{uni22}
\qquad \hspace{-1.0cm}
\mbox{for all} \ \eta\in V, \ \mbox{a.e.} \ t\in (0,T),\\[2mm]
&&\label{uni_init}
\hspace{-1.0cm}
u(0)=0, \ C(0)=0.
\end{eqnarray} 
Since $C\in L^2(0,T;V))$ equation \eqref{uni22} is equivalent to the following
\begin{eqnarray}
&&
\nonumber
\hspace{-1.0cm}
\langle C'(t)-d\Delta C(t) + B_1(u_1(t),C_1(t))-B_1(u_2(t),C_2(t)),\eta \rangle_{L^2(\Omega)} = \langle g\, C(t),\eta \rangle_{L^2(\Omega)} \\[2mm]
&&\label{uni2}
\qquad \hspace{-1.0cm}
\mbox{for all} \ \eta\in L^2(\Omega), \ \mbox{a.e.} \ t\in (0,T).\\[2mm]
&&\label{uni2a} \nonumber
\hspace{-1.0cm}
u(0)=0, \ C(0)=0.
\end{eqnarray}

\noindent 
First, observe that from Lemma~\ref{lemma1} we have for $u_1,u_2,u=u_1-u_2 \in E$
\begin{eqnarray}
&&\nonumber
a_1(u_1,u_1,u_1-u_2)-a_1(u_2,u_2,u_1-u_2)=a_1(u_1,u_1,-u_2)-a_1(u_2,u_2,u_1)
\\[2mm]
&&\label{a_1comp}
=a_1(u_1,u_2,u_1)-a_1(u_2,u_2,u_1) = a_1(u,u_2,u_1)=a_1(u,u_2,u).
\end{eqnarray}

\noindent
Moreover, from Lemma~\ref{Erhling} and $H(j)(e)$ we have
\begin{eqnarray}
&& \hspace{-1.0cm} \label{smallness_out}
-\langle \xi_1(t)-\xi_2(t),u_1(t)-u_2(t)\rangle \le m_1\|u(t)\|_{L^2(\Gamma)}^2 \le \varepsilon \|u(t)\|_E^2 + M(\varepsilon)\|u(t)\|_H^2
\end{eqnarray}
for $\varepsilon >0$ and a.e. $ t \in (0,T)$.
\noindent
Finally, choosing $v=u(t)$ and $\eta = -k\Delta C(t)$ 
in \eqref{uni1} and \eqref{uni2}, respectively and adding resulting equations gives, using $g\ge 0$,  \eqref{a_0b_0corecive}, \eqref{a_1comp} and  \eqref{smallness_out} we calculate
\begin{eqnarray} 
&&\nonumber
\hspace{-1.0cm}
\frac{1}{2}\frac{d}{dt}\Big(\|u(t)\|_H^2 + k\|\nabla C(t)\|_{L^2(\Omega)}^2 \Big) 
+ (\alpha-\varepsilon) \|u(t)\|_E^2 
+ k\, d\, \|\Delta C(t)\|_{L^2(\Omega)}^2  \\[2mm]
&&\label{uni3}\nonumber
\hspace{-1.0cm}
\quad \le -a_1(u(t),u_2(t),u(t)) + k(\langle -\Delta C_1(t)\nabla C_1(t),u_1(t)-u_2(t)\rangle \\[2mm]
&&\nonumber
+\langle \Delta C_2(t)\nabla C_2(t),u_1(t)-u_2(t)\rangle +\langle u_1(t)\nabla C_1(t),\Delta C_1(t)-\Delta C_2(t)\rangle\\[2mm]
&&\nonumber
-\langle u_2(t)\nabla C_2(t),\Delta C_1(t)-\Delta C_2(t)\rangle)  +M(\varepsilon) \|u(t)\|_H^2\\[2mm]
&&\nonumber
 \quad =  -a_1(u(t),u_2(t),u(t)) +k[\langle \Delta C_1(t)\nabla C_1(t),u_2(t)\rangle + \langle \Delta C_2(t)\nabla C_2(t), u_1(t)\rangle \\[2mm]
&&\nonumber
-\langle \Delta C_2(t)\nabla C_1(t)u_1(t)\rangle -\langle \Delta C_1(t)\nabla C_2(t),u_2(t) ] +M(\varepsilon) \|u(t)\|_H^2\\[2mm]
&&\nonumber 
= -a_1(u(t),u_2(t),u(t)) +k[-\langle \Delta C_2(t)\nabla C(t), u_1(t) \rangle +\langle \Delta C_1(t)\nabla C(t), u_2(t)\rangle \\[2mm]
&& \nonumber
+\langle \Delta C_1(t)\nabla C(t),u_1(t)\rangle -  \langle \Delta C_1(t)\nabla C(t),u_1(t)\rangle ] +M(\varepsilon) \|u(t)\|_H^2 \\[2mm]
&&\nonumber
=-a_1(u(t),u_2(t),u(t)) - k\langle \Delta C_1(t)\nabla C(t),u(t)\rangle +k\langle \Delta C(t)\nabla C(t),u_1(t)\rangle \\[2mm]
&&\nonumber
+ M(\varepsilon) \|u(t)\|_H^2
\end{eqnarray}
for $\varepsilon >0$ and a.e. $ t \in (0,T)$.
Hence, finally 
\begin{eqnarray}
&&\nonumber
\hspace{-1.0cm}
\frac{1}{2}\frac{d}{dt}\Big(\|u(t)\|_H^2 + k\|\nabla C(t)\|_{L^2(\Omega)}^2 \Big) 
+ (\alpha-\varepsilon) \|u(t)\|_E^2 
+ k\, d\, \|\Delta C(t)\|_{L^2(\Omega)}^2  \\[2mm]
&&
=-a_1(u(t),u_2(t),u(t)) - k\langle \Delta C_1(t)\nabla C(t),u(t)\rangle  \label{calculations_final}
+ k\langle \Delta C(t)\nabla C(t),u_1(t)\rangle 
\end{eqnarray}
for $\varepsilon >0$ and a.e. $ t \in (0,T)$.
We now estimate terms on the right-hand side 
of \eqref{calculations_final}. From Lemma~\ref{lemma1} and the Cauchy inequality with $\varepsilon >0$, 
we have
\begin{eqnarray}
&&\label{uni4}
\hspace{-1.0cm}
|a_1(u(t),u_2(t),u(t))|\le \varepsilon \, \|u(t)\|_E^2 + M(\varepsilon)\, \|u(t)\|_H^2\|u_2(t)\|_E^2,\\[2mm]
&&\nonumber
\hspace{-1.0cm}
|\langle \Delta C_1(t)\nabla C(t),u(t)\rangle| \le 
M\, \|\Delta C_1(t)\|_{L^2} \|\nabla C(t)\|_{L^4}\|u(t)\|_{L^4} \\[2mm]
&&\nonumber
\quad \hspace{-1.0cm}
\le \|\Delta C_1(t)\|_{L^2} \|\nabla C(t)\|_{L^2} ^{1/2} \| \Delta C(t)\|_{L^2} ^{1/2}\|u(t)\|_H ^{1/2} \|u(t)\|_E^{1/2} \\[2mm]
&&\label{uni5}
\qquad \hspace{-1.0cm}
\le \varepsilon \, (\|\Delta C(t)\|_{L^2}^2 + \|u(t)\|_E^2) +  M(\varepsilon) \, \| \Delta C_1(t)\|_{L^2}^2\|u(t)\|_H^2\|\nabla C(t)\|_{L^2}^2, \\[2mm]
&&\label{uni6}\nonumber
\hspace{-1.0cm}
|\langle \Delta C(t)\nabla C(t), u_1(t)\rangle| \le 
M\|\Delta C(t)\|_{L^2}^{3/2}\|\nabla C(t)\|_{L^2}^{1/2}\|u_1(t)\|_{L^4}
\\[2mm]
&& \label{uni7} 
\qquad \le \varepsilon \, 
\|\Delta C(t)\|_{L^2}^2  + M(\varepsilon) \, 
\|\nabla C(t)\|^2\|u_1(t)\|_H^2\|u_1(t)\|_E^2
\end{eqnarray}
for a.e. $t\in (0,T)$.
\noindent 
We now choose 
$\varepsilon <\min\{\frac{\alpha}{3},\frac{kd}{2}\}$.
From \eqref{uni4}--\eqref{uni6} applied to the right hand side
of \eqref{uni3}, we get
\begin{eqnarray}
&&\nonumber
\hspace{-0.7cm}
\frac{1}{2}\frac{d}{dt}\Big(\|u(t)\|_H^2 + k\|\nabla C(t)\|_{L^2(\Omega)}^2 \Big) + (\alpha-3\varepsilon)\|u(t)\|_E^2 + (kd-2\varepsilon)\|\Delta C(t)\|_{L^2}^2 \\[2mm]
&&
\qquad \nonumber
\le M \,(\|u(t)\|_H^2\|u_2(t)\|_E^2 + \|\Delta C_1(t)\|_{L^2}^2\|u(t)\|_H^2\|\nabla C(t)\|_{L^2}^2 \\[2mm]
&&\qquad\nonumber 
+ \|\nabla C(t)\|^2\|u_1(t)\|_H^2\|u_1(t)\|_E^2 +  M(\varepsilon) \|u(t)\|_H^2 \\[2mm]
&& \hspace{-1.8cm}
\qquad \le M(\|u(t)\|_H^2 + \|\nabla C(t)\|_{L^2}^2)(\|u_2(t)\|_E^2 + k\|\Delta C_1(t)\|_{L^2}^2 + \|u_1(t)\|_H^2\|u_1(t)\|_E^2 +1) \label{last}
\end{eqnarray}
for a.e. $t\in (0,T)$.
\noindent 
By estimates \eqref{bound_uinf}--\eqref{bound_CH2}, 
it is clear that 
functions $u_1$ and $u_2$ belong to
$L^2(0,T;E)\cap L^\infty(0,T;H)$, 
and $C_1$ belongs to
$L^\infty(0,T;H^1(\Omega))\cap L^2(0,T;V)$. 
Integrating \eqref{last} over $(0, t)$ for $t \in (0, T)$ 
and applying the Gronwall lemma, we obtain 
\begin{equation}\label{unilast}
\|u(t)\|_H^2 + k\, \|\nabla C(t)\|_{L^2(\Omega)}^2 \le 
M \, (\|u(0)\|^2 + \|C(0)\|^2) 
\end{equation}
for a.e. $t\in (0,T)$.
\noindent 
Finally, from conditions~\eqref{uni_init} and
% we have $u(0)=0$, and $C(0)=0$, hence we conclude from
\eqref{unilast}, we conclude that $u\equiv0$ and $C\equiv0$. This proves the uniqueness of solution to Problem~\ref{PVar}.
$\hfill{\square}$

\medskip

%\begin{Remark}
	In this paper we have studied a two-dimensional fluid flow and we have left three-dimensional problem for a future work since this problem would be more difficult and the solution would be probably defined only on a smaller time interval. Moreover, it would be interesting to consider in the future work a problem not with the slip boundary condition, but with a leak boundary condition, see \cite{KASHI} for Navier-Stokes problem with the leak condition.
%\end{Remark}

\end{document}